\documentclass{ws-ijmpe}
\usepackage[super,compress]{cite}

\def\be{\begin{eqnarray}}\def\ee{\end{eqnarray}}
\def\lsim{\mathrel{\rlap{\lower3pt\hbox{\hskip1pt$\sim$}}
     \raise1pt\hbox{$<$}}} 
\def\gsim{\mathrel{\rlap{\lower3pt\hbox{\hskip1pt$\sim$}}
     \raise1pt\hbox{$>$}}} 
\def\le{ \begin{array}{ll}}\def\re{\end{array}}

\def\lear{ \left( \begin{array}{cc}}\def\rear{\end{array} \right)}

\def\le{ \left( \begin{array}{cc}}\def\re{\end{array} \right)}

\def\bi{\bibitem}

\def\la{\langle}\def\ra{\rangle}

\begin{document}
\title{Going from Asymmetric Nuclei to Neutron Stars
\\ to Tidal Polarizability in Gravitational Waves}

\author{Mannque Rho}

\address{Institut de Physique Th\'eorique, CEA Saclay, 91191 Gif-sur-Yvette c\'edex, France\\
E-mail: mannque.rho@ipht.fr}
\author{Yong-Liang Ma}

\address{Center for Theoretical Physics and College of Physics, Jilin University, Changchun,\\
E-mail: yongliangma@jlu.edu.cn}

\maketitle

\begin{abstract}
Among the five-year government-funded World Class University Projects in Korea, the category-3 program approved at Hanyang University in Seoul led to an exploratory effort to go from neutron-rich nuclei to dense matter in neutron stars.   The principal results in what transpired in the effort -- and what followed afterwards --  are described with the focus on the possibly important, hitherto unexplored, role  played in nuclear dynamics of topology and hidden symmetries of QCD. The potential link to the proton mass problem is pointed out.
\end{abstract}



\section{Objective}
The Korean Government launched in 2008 a 5-year national project entitled  ``World Class University (WCU)," the official objective of which was to elevate  the Korean university level of both science and humanities to the forefront of the world. Though not officially stated, this was a prelude to the subsequent establishment of the ambitious multi-billion dollar research center that comes under the name of ``Institute for Basic Science (IBS)," with some two-third of the funding allotted to go to a high-quality RIB accelerator devoted to nuclear science. Among the three classes of the WCU program, there was the third program  labelled as ``WCU-III" that was focused solely on basic research, with renowned scholars invited from abroad to participate for a few months per year in joint research activities.   The WCU/Hanyang,  directed by Hyun Kyu Lee of the physics department, was of this category. We both participated closely in this project, MR as an invited scholar and YL as a foreign collaborator.  The WCU/Hanyang program (W/H for short), in anticipation of the forth-coming IBS with the RIB machine playing the frontier physics tool, had the implicit objective of exploring what could be the goal of the accelerator in project, known under the name of  ``RAON,"  for what are considered as fundamental issues in nuclear physics. The underlying theme at the WCU/Hanyang was  the problem of unraveling the origin of the proton mass,  arguably a ``mass without masses"~\cite{wilczek} a mystery unresolved even after the discovery of Higgs boson. The key idea in the work was  to ``undress" or ``melt" the proton mass by high density, hence a nuclear physics problem, and high density inevitably requires compression by gravity, i.e.,  compact stars, hence an astrophysics problem.

The objective of the W/H  was not directed specifically toward confronting what is to be measured when the machine is first turned on. This task was to be fully  taken up by  a large number of researchers working on exotic nuclei to be explored. Our entire aim was  to go from what can be learned at RAON  on, say,  highly asymmetric nuclei to the extreme regime relevant to neutron stars with high density. So, the question raised was how what is studied  in largely asymmetric nuclei at RAON can be extrapolated to decipher what may be taking  place in massive compact stars and in what is observed from coalescing massive compact stars. We believed that the answer to this question also answers where the proton mass comes from.

The compact-star matter involves a high density regime that can be accessed neither by perturbative QCD  nor by lattice QCD, the only nonperturbative model-independent tool known. Thus in the absence of laboratory experiments, it  is as yet a totally uncharted domain. Therefore, to make {\it any} progress at all,  we were led to the conclusion that certain guesses and/or assumptions and some sort of speculations are inevitably involved.

In this article we describe some of the key results obtained in the W/H  and the progress made since the  W/H~\cite{cnd3}.  We must admit that although some results are in fairly good  agreement with Nature, some others, mostly in the form of novel  predictions, await to be vindicated or possibly ruled out by observations.  Many of the issues are currently highly controversial and subject to debate. This paper highlights some features that are not widely familiar to many physicists in the field, both theoretical and experimental, which we believe, if correct, will impact on the development in the future.

Some 400 references are involved. They can be found in the monograph~\cite{cnd3}. Here we will cite only a few.

\section{Effective field theory approach to QCD}
The key approach adopted in the W/H program -- and continued after  -- was anchored on, with some perhaps daring or unjustifiable extrapolations of,  Weinberg's ``Folk Theorem (FT)" for quantum effective field theories~\cite{weinberg}. Quoting Weinberg,  it states ``If one writes down the most general possible Lagrangian, including all terms consistent with assumed symmetry principles, and then calculates matrix elements with this Lagrangian to any given order of perturbation theory, the result will simply be the most general possible S-matrix consistent with perturbative unitarity,  analyticity, cluster decomposition, and the assumed symmetry  properties." The attitude we are taking is ``how can this theorem go wrong?"

We think it is safe to say that what has taken place in nuclear theory since 1990's,  which can claim some success, amounts to implementing as faithfully as feasible this Folk Theorem in nuclear dynamics. Thus it must be that the more faithful to the premises for the Folk Theorem -- and the harder and astutely one works in going to higher orders, the closer the theory will approach Nature. It is along this line of reasoning that the WCU/Hanyang program proceeded.

In this paper we present briefly the highlights of this development

There are three classes of phenomena where the workings of the FT show up differently: (1) precision calculations, (2) global structure of normal nuclear matter and (3) extrapolation to high density.  RAON physics is to address the first two and compact-star physics to the third. The FT is in some sense confirmed in the first two classes, so what's to be done there is figuring out how fully the requirements for the FT can be implemented. For the third, on the contrary, one is bound to go beyond what's known or established and hence highly explorative.
\subsection{Relevant degrees of freedom}
The strategy adopted in the W/H was  to adhere  to the FT using a {\it one and only one} effective field theory Lagrangian, expressed {\it entirely} in terms of hadrons in  addressing  issues ranging from RAON-type physics to massive compact stars.  The degrees of freedom taken into consideration consist of  the nucleons (or more generally light-quark baryons including $\Delta$ resonances and hyperons), the pseudo-Nambu-Goldstone bosons (``pions" in $N_f=2$ or 3 flavors), a  pseudo-Nambu-Goldstone scalar  (called dilaton), and the lightest vector mesons, $\rho$ and $\omega$, introduced as hidden gauge fields\footnote{Infinite towers of vector mesons as in gauge-gravity duality models do contribute to certain quantities. This feature was treated in the W/H and in the monograph.  Here we skip that matter.}. Let us refer to the effective Lagrangian so constructed -- and matched to QCD at a matching scale -- as ``EFT-$bs$HLS" with $b$ standing for baryons, $s$ standing for scale symmetry and HLS standing for  hidden local symmetry gauge-equivalent to non-linear sigma model.\footnote{Later ``EFT" in EFT-$bs$HLS will be specified as ``$V_{\rm lowk}$RG"  explained below. Unless mentioned otherwise,  it will be denoted simply as $bs$HLS for short.} In the W/H program, baryons were also generated as skyrmions from the Lagrangian without baryons explicitly incorporated. The corresponding Lagrangian will be denoted $s$HLS.  There is  power present in the topological approach that is not apparent in $bs$HLS,  and that provides a crucial input in $bs$HLS as  explained below.

\subsection{Hidden symmetries}
Hidden symmetries play the key roles in $bs$HLS.
Apart from chiral symmetry which is fairly well understood in nuclear dynamics, there are certain symmetries which are ``hidden" -- in the sense that they are not explicitly visible -- in QCD. There may be several of them such as Weinberg's ``mended symmetries" but here we focus on two.  One is scale symmetry, which is broken by the trace anomaly (and quark mass), and local flavor symmetry, which is broken by the visible mass much greater than those of pseudo-Nambu-Goldstone bosons. The first applies to a scalar meson $\sigma$ and the latter to the vector mesons $V=(\rho, \omega)$. In what one calls ``standard" chiral perturbative approach (S$\chi$PT for short)  on which ``standard" nuclear effective field theory (S$\chi$EFT)  is anchored, both $\sigma$ and $V$ do not figure explicitly. They are assumed to be simulated at higher order chiral perturbation.

In the W/H program and in the subsequent development summarized in \cite{cnd3},  both scale symmetry and local (gauge) symmetry are {\it explicitly} incorporated in the nonlinear sigma model that gives rise to  S$\chi$PT. The former figures as a (pseudo-)NG boson dilaton identified with the scalar $f_0(500)$  listed in the particle data booklet. It is assumed to have an assumed infrared (IR) fixed  point for three-flavor systems~\cite{ct}. This is currently highly controversial, requiring further debates.  The latter are flavor gauge boson fields, hidden in QCD,  which could become visible at what's known as the vector manifestation (VM) fixed point at which the flavor gauge coupling $g$ goes to zero and the meson becomes massless~\cite{HY}. The basic idea developed in the W/H is that both can emerge in some sense and play an extremely important role in dense medium. This aspect  is basically different from  S$\chi$EFT. Some of the issues involved are highly controversial with lots of disputes, both pro and con, among workers in other area of physics, e.g., dilatonic  Higgs boson and beyond the Standard Model. They are also highly controversial in nuclear processes. Nonetheless it is found in the W/H program that they can be highly relevant and play a  key role in going beyond normal nuclear matter regime  to high density regime.

\subsection{Topology change}
\subsubsection{Hadron-quark continuity}\label{hadron-quark}
At some high density above the normal nuclear matter density $n_0\approx 0.16$fm$^{-3}$, baryonic matter is expected to have quarks and gluons appearing, explicitly or mixed with hadrons, in the system. At what density this can happen is not known.  A rough guess puts it at  $\gsim 2n_0$. Although it is not rigorously shown, it seems that strongly coupled quarks should figure at $n\gsim 2n_0$,  in a continuous transition from baryons to quarks, in order to accommodate the observed $\sim 2$-solar mass star~\cite{baymetal}. If this is what happens in nature, the question then is: How can the Lagrangian adopted in the W/H program -- which has only hadrons as relevant degrees of freedom -- account for the hadron-quark continuity? The  answer found in the W/H was that when baryonic matter is described in terms of skyrmions -- which is a valid description in the large $N_c$ limit in QCD, the skyrmions fractionize into half-skyrmions at some high density that we denote as $n_{1/2}$.  This process is topological, so could be considered as robust, as in condensed matter physics. At what density this takes place cannot be estimated by theory, but within the EFT-$bs$HLS formalism, it must happen not far above $n_0$. 

The key strategy in the W/H is to ``translate" this topology change into the ``bare" parameters of the $bs$HLS. We identify it as what corresponds to a sort of hadron-quark continuity in QCD variables. It is tantamount to trading in quarks/gluons  for topology. This is along the line of the skyrmion-quark trading  known as ``Cheshire-Cat phenomenon" found in 1980's~\cite{CC}. As is discussed extensively in \cite{cnd3}, this Cheshire-Cat property is ubiquitous at low as well as high densities -- in fact all the way to the color-flavor locking at asymptotic density. An attractive way of formulating such ``trading" is Holger Bech Nielsen and his collaborators' idea of a gauge symmetry for confinement-deconfinement process at low energy.
\subsubsection{Nuclear tensor forces}\label{tensor forces}
The topology change entails a number of surprising phenomena, not at all obvious in S$\chi$EFT.
It brings in the half-skyrmion phase and connects to the vector manifestation (VM) behavior in the gauge coupling $g$ to the $\rho$ meson in the $bs$HLS, which vanishes at the critical density at which chiral symmetry is restored $n_{vm}\gsim 20 n_0$. This causes the tensor force, which drops as density increases, to abruptly change over and increase past $n_{1/2}$.
{\it This makes at $n_{1/2}$ a cusp in the symmetry energy which crucially controls the equation of state (EoS) of neutron-rich matter~\cite{cusp}.}

This cusp structure in the symmetry energy was first {\it predicted}  in the skyrmion matter in the leading order in $1/N_c$ in the skyrmion quantization. It is topological. This represents the principal impact of topology on the EoS which will be manifested in the change of degrees of freedom associated with an emerging scale symmetry, i.e., pseudo-conformality.
\subsubsection{Parity doubling} \label{parity doubling}
The topological consideration reveals another symmetry property hidden in continuum Lagrangians. In skyrmion descriptions, the bilinear quark condensate  $\la\bar{q}q\ra$, which is conventionally considered as an order parameter of chiral symmetry, vanishes when averaged over space, $\Sigma\equiv \overline{\la\bar{q}q\ra}\to 0$. This, contrary to what one might naively think,  does not signal chiral symmetry restoration because the pions are still present in the half-skyrmion medium. This state with the vanishing $\Sigma$ is somewhat like the pseudo-gap phase in high-T superconductivity where the surperconductivity gap is nonzero whereas the paring is zero. An important consequence is that this causes parity doubling  in the half-skyrmion phase, with the  effective baryon mass (skyrmion mass)  going to a nonzero value, $m_0$,  as density approaches chiral symmetry restoration density.  This (chiral scalar) mass $m_0$ turns out to be substantial, $m_0\sim (0.6-0.9) m_N$, and signals the origin of the nucleon mass~\cite{protonmass}, throwing doubt to the lore that the proton mass is, mostly if not entirely, generated from the vacuum fluctuation, i.e., by the spontaneous breaking of chiral symmetry. We will see below that this parity doubling has a dramatic consequence on the symmetry property of compact stars.

\subsection{Matching to QCD}\label{matching}
An effective field theory that is defined at a suitable energy scale needs to be matched to  QCD to be ``faithful" to the FT. How this matching which is done via correlators is effectuated  is important for the W/H where the notion of ``intrinsic density dependence (IDD)" figures importantly.

The matching scale is typically the chiral scale $\Lambda_\chi=4\pi f_\pi\sim 1$ GeV where $f_\pi \simeq 93$ MeV is the pion decay constant. In practice, the matching is done at a scale lower than $\Lambda_\chi$. In nuclear physics, one usually works with a cutoff slightly below the vector-meson mass as is done in S$\chi$PT. In the framework developed in the W/H, the matching scale (denoted as $\Lambda_M$)  is taken somewhat above the vector meson mass,  given that the vector mesons figure as the explicit degrees of freedom.  What the matching does is then to define the EFT Lagrangian with the ``bare" parameters endowed with nonperturbative QCD information, e.g., such vacuum condensates  as quark, gluon, dilaton etc.  Since the condensates represent the properties of the vacuum, when the vacuum is changed by density as in our case, the condensates ``slide" as density varies. This endows the EFT-$bs$HLS with sliding density dependence, namely, the IDD mentioned above.  As defined, the IDD is a Lorentz-invariant quantity.

In practical calculations, the effective cutoff is usually taken at a value $\Lambda^{\rm eff}$ lower than $\Lambda_M$. In the renormalization-group sense, the effect of decimating from $\Lambda_M$ to $\Lambda^{\rm eff}$ entails changes to the ``bare" parameters of the EFT Lagrangian. The density dependence brought in by this effect is referred to as ``induced density dependence," DD$_{\rm induced}$ for short. This dependence is  Lorentz non-invariant.  Whenever necessary, we will denote in what follows the sum of IDD and  DD$_{\rm induced}$ as IDD$^*$. In applications, we do not make, unless otherwise stated,  the distinction between IDD and IDD$^*$ because that IDD$^{\rm induced}$ is usually a small correction that one can ignore in semi-quantitative estimates.

We should note that in the S$\chi$PT approach to nuclear physics, the Lorentz-invariant IDD is absent.  There the bare parameters are fixed at the vacuum values. Some parts of DD$_{\rm induced}$ do, however, figure there. The example is the effect of three-body force in effective two-body forces in the standard approach.

\section{How the FT fares in nuclear processes}
At low densities near the equilibrium nuclear matter density $n_0$, one can make, in faithful implementation of the Folk Theorem,  a precision calculation with theoretical errors much less than $1\%$~\cite{precision}.  In some cases like the solar HEP process, one can eliminate orders-of-magnitude uncertainties that have been plaguing theoretical efforts for a long time. Up to $n_0$,  S$\chi$EFT and the W/H's EFT-$bs$HLS are comparable in their quality of fits to nature. However  at higher densities above $n_0$, particularly going beyond the topology change density $n_{1/2}$, the former loses its predictive power,  whereas  explorative calculations in the latter can be done without being bogged down with numerous unknown parameters and conceptual deficiencies.
\subsection{Chiral filter mechanism}\label{chiral filter}
In order to see how the FT renders  precision calculations feasible in nuclear physics, it is extremely helpful to see how the soft-pion theorems developed in 1970's in connection with the current algebras figure in the processes. This was dubbed then as ``chiral filter mechanism"~\cite{kdr}.  In the modern chiral effective theories anchored on nonlinear sigma model for low-energy hadronic interactions, the soft-pion theorems are encoded in chiral Lagrangians at the tree level.  So they are included  automatically when one works out chiral expansion order by order.

However there is something magical about the soft-pion theorems, in the same way as in other soft theorems such as soft gravitons etc, being discussed in other areas of physics.

In the approach predominantly adopted by nuclear theorists,  one uses nuclear forces and wave functions. In fact the Weinberg scheme of S$\chi$EFT relies on them expanded in terms of irreducible and reducible diagrams. In this approach,  soft pions are found to play a particularly prominent role in certain channels in nuclear electroweak response functions. In the systematic chiral expansion of nuclear electro-weak currents, it is in the exchange currents, specially two-body ones, where soft pions figure very importantly. In fact it was known since late 1970's, before the advent of QCD,  that two-body exchange currents for magnetic dipole and weak axial charge transitions give huge model-independent corrections to the leading singe-particle operators, with higher-order corrections strongly suppressed -- by more than chiral two orders. Thus the nuclear axial charge and magnetic dipole transitions can be very accurately calculated if the wave functions are accurately computed. The well-known examples are the thermal neutron-proton capture $n+p\to \gamma +d$ and  the first-forbidden axial-charge transitions $A(0^\pm)\leftrightarrow B(0^\mp)$ with $\Delta T=1$ accompanied by a neutrino (or anti-neutrino). This soft-pion dominance has been beautifully confirmed in processes involving light nuclei where accurate wave functions are available. Certain processes are computed with a theoretical uncertainty less than 1\%, which validates the FT to a great accuracy. One can say those  processes are protected by ``chiral-filter."

A corollary to this chiral filter protection applies to the electric charge operator and the weak Gamow-Teller operator. Soft pions are suppressed in these operators, so the exchange current contributions  are strongly suppressed -- by as much as three chiral orders (denoted in the literature as N$^3$LO) -- relative to the leading single-particle operator. Thus if the corrections which enter at N$^3$LO is negligible, then one can simply ignore the corrections. But if the $N^3$LO corrections are not negligible relative to the leading term -- which can happen when the single-particle operator is accidentally suppressed by symmetry or the process involves large soft momentum transfer as in the case of the Gamow-Teller matrix elements in neutrinoless double beta decays --  then one has to go much beyond N$^3$LO.
 In this case, the chiral expansion has no predictive power since the series could converge badly and the number of  un-determinable constants typically blows up.  It is unfeasible to obtain reliable corrections by going to only a few manageable orders in the chiral counting. As stressed in the requirements for the FT, ``softness" is the key to the success of effective field theories anchored on chiral symmetry.  This of course does not mean that exchange current corrections to these chiral-filter-unprotected processes are necessarily negligible.  It would require a different organization of terms than the standard chiral ordering. We illustrate  this feature with the so-called ``quenched $g_A$ problem" in nuclear Gamow-Teller strength, a long-standing mystery.
\subsection{Nuclear effective chiral field theory vs. energy density functionals} \label{eft-edf}
Although we did not dwell on the issue in the W/H, we should  mention that there is an actively on-going effort to approach nuclear structure and nuclear EoS, approaching many-body problems in powerful numerical techniques. This approaches the FT with the nuclear potentials and electroweak currents in a variety of S$\chi$PT schemes treated at N$^k$LO (with $k$ typically 3 or 4). Beyond $k=4$ as needed for the Gamow-Teller matrix elements in double beta decays, however, the problem is hampered by the lack of data to control the growing number of problems and becomes powerless.

While different in spirits, the variety of energy density functional approaches do a similar effort toward meeting the requirements for the FT.  The renormalization-group approach, EFT-$bs$HLS, as applied to compact stars, belongs to this class This formulation can be related to a Landau Fermi-liquid approach.
\subsection{Tensor force in action}
\subsubsection{C14 dating}
One of the most prominent impacts of the nuclear tensor forces discussed in Section  \ref{tensor forces} is the effect of the strength of the net tensor force for $r\gsim 0.7$ fm that decreases as density approaches $n_{1/2}$, producing the cusp in the symmetry energy. The effective tensor force nearly vanishes at $n\sim 2n_0$. This is spectacularly manifested in the long half-lifetime of C-14, as discovered in a shell-model calculation by Jeremy Holt and his Stony Brook collaborators~\cite{c14}. The Gamow-Teller matrix element responsible for the decay is fine-tuned to nearly vanish near $n_0$  by the diminishing tensor force with no effect due to meson-exchange currents. That there is no appreciable exchange current effect is consistent with the chiral filter mechanism discussed above. We will come back to this issue in connection with the problem  of quenched $g_A$.
\subsubsection{In exotic nuclei}
We now come to the issue that belongs to the physics that the RAON machine can address. For this, we need a Wilsonian renormalization-group idea applied to the $V_{\rm lowk}$ strategy starting with the $bs$HLS Lagrangian introduced above. The $V_{\rm lowk}$ developed at Stony Brook is the potential obtained from the ``bare" potential with the ``bare" parameters of the EFT Lagrangian determined at the scale $\Lambda^{\rm eff}$. It is obtained by the Wilsonian renormalization decimation by fitting the whole  set of experimental data available up to the lab energy commensurate with the decimation cutoff.  In the way formulated, it should be a fixed point quantity that is more or less independent of the cutoff. It has the merit to be free of the short-distance physics taking place above the cutoff scale.

Renormalization decimations can be performed in several steps. In the W/H, two decimations were adopted. Given the $bs$HLS Lagrangian, one can perform either single decimation or double decimations. The single decimation corresponds to the Landau Fermi-liquid fixed point theory with $1/\bar{N}$ corrections (where $\bar{N}=k_F/\bar{\Lambda}$ with $\bar{\Lambda}$ the cutoff relative to the Fermi surface) suppressed. The class of relativistic mean-field theories popular in the field belongs to this approach, with the crucial difference from our approach being the absence of IDDs (while DD$_{\rm induced}$ is at least partially included by higher field corrections usually made  which leads to an improved compression modulus).

The second decimation in the double-decimation approach implements higher order $1/\bar{N}$ corrections, which accounts for fluctuations on top of the Fermi-liquid fixed point. The Stony Brook $V_{\rm lowk}$ renormalization group approach ($V_{\rm lowk}$RG for short) is adapted to doing -- via ring diagrams --  this double-decimation procedure.   In applications to compact-star matter, both the single decimation with $bs$HLS and the double decimation with $V_{\rm lowk}$, i.e., $V_{\rm lowk}$RG were applied.

%
The monopole matrix element that we describe here in $V_{\rm lowk}$RG was not the main focus of the W/H activity, but it underlies the basic premise of the approach that illustrates the uncanny role of the tensor force in RAON-type physics. It has to do with the role of the nuclear tensor force in the shell evolution in exotic nuclei,  based principally on the series of work done by T. Otsuka and his collaborators at the University of Tokyo\cite{otsuka}.  Let's refer to this work as UT.

The key observation made by the UT is  that the monopole matrix element (MME) of the single-particle state is dominated by the tensor force which can be neatly extracted from other forces This problem can be treated equally well with sophisticated phenomenological potentials, but we phrase the problem in terms of the ``bare" potential in $bs$HLS Lagrangian. This is because it may offer a means to probe  the IDDs and the vacuum structure sliding with density.

In $bs$HLS, the net tensor force is  the sum of one-pion exchange  and one-$\rho$ exchange. With the signs opposite,  the {\it net} tensor force is reduced from the pion tensor in the relevant range of interaction for $r\gsim 0.7$ fm. The UT considered three possible scenarios but ignored the possible role of IDDs. They first compute the MME with the ``bare" tensor force, second with the $V_{\rm lowk}$ obtained by the RG decimation in the free-space and third the $V_{\rm lowk}$RG  in the double decimation (i.e.,  with a selected  $1/\bar{N}$ corrections taken into account).  It was found that all three gave (nearly) the same tensor contribution to the MME. This indicated there is ``no renormalization" both in the free space and in medium. Put in the language of renormalization group, this states that the $\beta$ function for the tensor force is equal to zero,  meaning that  the tensor force is a fixed-point quantity.  The tensor force as used in the UT work is  finite-ranged, so it is nonlocal.  In effective field theory approaches to Fermi-liquid systems, in particular the Fermi-liquid fixed point theory, one limits to local four-Fermi interactions. It is unclear what the fixed-point structure of the nonlocal tensor force is. But it does come out that $V^T_{\rm lowk}\approx V^T_{\rm bare}$.

 In the UT work, the ``bare" parameters in the Lagrangian, say, our $bs$HLS did not contain IDDs. That the description worked well without IDDs means either that IDDs are absent or that the process did not prove the sliding vacuum effect. The C14 dating process  exhibited an intricate connection between the effective density regime probed by the wave functions, the energy splitting and the Gamow-Teller matrix element.  Studying whether and how the density dependence of the tensor force manifests in the shell evolution would be an interesting problem. Given that the MME  indicates non-renormalization in the tensor force both in the free space and in medium, that is, QCD strong interactions, it would be a pristine signal for the possible IDDs manifested in nuclei in a way different for what's observed in the C14 dating.  The crucial question could be: Is this apparent fixed-point property of the tensor force dependent on specific channels or globally for the tensor force as a whole? If it were for the latter, it would have an extremely interesting implication in what happens in compact-star matter to which we turn below.

\section{Emergence of hidden symmetries}
Up to the normal density and slightly above, perhaps up to the putative topology change density $n_{12/}$, effective field theories based on S$\chi$PT should be more or less successful in describing  nature. The W/H approach in $V_{\rm lowk}$RG would do just as well in that density regime, with no great difference from the standard approach. The degrees of freedom associated with hidden symmetries did not seem to figure explicitly. However what transpired from the W/H is that  a significant difference can come in at high densities $n\gsim n_{1/2}$  due to the  symmetries that are not visible at low densities.
\subsection{Hidden local symmetry}\label{hls}
Up to $n_0$ and most likely up to $\sim n_{1/2}$, flavor local symmetry is hidden, presumably buried in higher chiral-order terms in S$\chi$PT.  Beyond $\sim n_{1/2}$, the hidden local symmetry encodes the vector manifestation that is absent, at least at low orders,  in S$\chi$PT. It has been shown in an RG analysis of HLS Lagrangian that as the quark condensate drops to zero as chiral restoration is approached, the  gauge coupling $g$ goes to zero following the quark condensate that goes to zero,  $g\sim \la\bar{q}q\ra\to 0$. At that point  the gauge symmetry is to emerge, with massless vector-meson excitations, in particular the $\rho$.  Whether or not this theoretical prediction does take place in nature is not yet settled. However we see a precursor to this property in the appearance of the cusp in the symmetry energy with the changeover in nature of the tensor force mentioned above.  This precursor behavior will be found to have a striking consequence in what follows for $n\gsim n_{1/2}$.
\subsection{Hidden scale symmetry}\label{hss}
The current theoretical activity on what is referred to as the ``first-principle approach to  nuclear effective field theories" anchored on S$\chi$PT scheme, i.e,  S$\chi$EFT, has no explicit scalar degree of freedom, and hence no reference to scale invariance broken quantum mechanically in QCD.  Scalar excitations could be present only if they were generated at high orders in the chiral expansion. The W/H approach argues that this S$\chi$EFT must break down as density exceeds $n_{1/2}$.


Even in the density regime where  S$\chi$PT is applicable, there is power in implementing scale symmetry {\it ab initio}. The reason  is that scale invariance is in fact present in the nonlinear sigma model on which S$\chi$PT is based. It is just  hidden~\cite{ct,yamawaki}. This statement follows from the reasoning provided by K. Yamawaki~\cite{yamawaki}.  Starting with a version of linear sigma model to which the Nambu-Jona-Lasinio model is equivalent, one can do a series of field redefinitions to rewrite the Lagrangian in a form in which there is a parameter available, say, $\lambda$,  that {\it can be tweaked}. If $\lambda$ is dialed to the strong coupling limit, the Lagrangian goes to the nonlinear sigma model to which hidden local symmetry is gauge equivalent, hence embodying low-energy dynamics. If, on the other hand, one dials to  the weak coupling limit, then one arrives at a scale-invariant Lagrangian, modulo a potential that contains explicit scale symmetry breaking. Thus what matters is the dialing of $\lambda$.  {\it The claim made in the W/H work is that the density does the dialing of the $\lambda$.}

This notion will be found to have an interesting nontrivial prediction at the weak coupling regime at high density. But it also turns out  to solve a long standing problem in nuclear physics in the strong coupling regime, i.e., the  ``quenched $g_A$".
\subsubsection{$g_A$: To quench or not to quench, that is the question}
The answer to this question is ``not to quench"~\cite{gA}.

And this question has a long history dating from early 1970's~\cite{wilkinson}.

When the Gamow-Teller matrix element for zero momentum transfer is calculated in simple shell model\footnote{By simple, we mean one open-shell, corresponding to the Fermi surface in Landau Fermi-liquid theory.} in light nuclei, it has been found that the effective axial-vector coupling constant $g_A^{\rm eff}$ needs to be quenched by about 20\% to explain the experimental data. This observation, which still holds nowadays and also in Gamow-Teller resonances, has spurned a long-standing debate since 1970's when Denys Wilkinson (and others) started to seriously investigate the problem, both experimentally and theoretically. Quite remarkably, the quenched constant is established to be close to $g_A^{\rm eff}=1$, with only a small variation due to both experimental and theoretical uncertainties. Since one expects that $g_A$ should go  {\it exactly} to 1 in the Wigner-Weyl phase where chiral symmetry is restored, a natural question was whether the observation $g_A^{\rm eff}\approx 1$ in nuclei is a precursor to the restoration of spontaneously broken chiral symmetry predicted at some high density in QCD. This question is quite  natural  since it is more or less accepted that the pion decay constant $f_\pi$, closely related to $g_A$,  should scale down in nuclear medium as density increases and go to zero at some high density. Theoretically $f_\pi$ does decrease as the quark condensate $\la\bar{q}q\ra$ does at increasing density.

So does $g_A$ get affected by the decreasing quark condensate?

The answer to this question is found in the role hidden scale symmetry and hidden local symmetry combined play in low-density systems with $n < n_{1/2}$. When scale symmetry is implemented as described in Section \ref{hss}, the pion decay constant in medium is accompanied by the ratio of the dilaton condensates $\Phi=\la\sigma\ra^\ast/\la\sigma\ra_0$ -- corresponding to the IDD defined in Section \ref{matching} -- where the superscript $\ast$ stands for density dependence and the subscript $0$ stands for the vacuum, as $f_\pi^\ast=\Phi f_\pi$. The light-quark hadron masses also slide down similarly. In stark contrast,   however, $g_A$ is scale-invariant and remains un-sliding with density, $g_A^\ast=g_A$. This means that the effect that makes $f_\pi^\ast$ decrease as the quark condensate drops does not affect  the Gamow-Teller coupling constant. This means $g_A^{\rm eff}\to 1$ must have nothing to do with a possible precursor to $g_A=1$ in the chiral-symmetry restored phase. The simple conclusion is that $g_A^{\rm eff}\to1$ must be due purely to nuclear correlations unrelated directly with IDDs.

This can be easily seen in Landau-Migdal Fermi-liquid fixed point theory.\footnote{Migdal figures importantly in extending Landau theory to nuclear dynamics. This is a natural extension, given that the equilibrium nuclear matter can be understood as a matter at a Fermi-liquid fixed point.}

The extremely simple solution is gotten at large $N_c$ and large $\bar{N}=k_f/\bar{\Lambda}$. In QCD, $g_A$ is considered to go as $O(N_c)$, so the large $N_c$ limit is appropriate here. Now in the nuclear EFT-$bs$HLS, one may also apply the large $\bar{N}=k_F/\bar{\Lambda}$ limit. In these two limits, the Landau mass for the quasiparticle, $m_L$, can be related to the fixed-point quantity $g_A^L$ by $m_L/m_N=\Phi \sqrt{g_A^L/g_A}$.  From the single-decimation calculation with $bs$HLS, one finds $m_L$ given in terms of the Landau parameter $F$, so $g_A^L/g_A$ is given once $\Phi$ is fixed. The $\Phi$ determined from the value of $f_\pi^\ast$ in deeply bound pionic systems yields  $g_A^L/g_A\approx 0.79 $ {\it nearly independently of density} between $n_0/2$ and $n_0$. This therefore gives  $g_A^L=0.79 \times 1.27\approx 1.00$ for the currently accepted value $g_A=1.27$. This way of resolving the problem  relies on the assumption that the $\Delta$ resonances -- and also other high-lying baryons -- can be integrated out. Including them explicitly in the Fermi-liquid approach is problematic as argued below.

Now what does $g_A^L\approx 1.00$ mean for the ``observed" quenched $g_A^{\rm eff}$?

There are three relevant issues involved here:
\begin{enumerate}
\item If the large $N_c$ consideration is trustworthy, then $g_A^L$ is the effective coupling constant for the quasiparticle on the Fermi surface to make the Gamow-Teller transition with zero momentum transfer.  This is the full transition matrix element in Landau-Migdal Fermi liquid fixed-point theory, going beyond the mean-field approximation  in the large $\bar{N}$ limit.  It is obtained for infinite matter. But given that  it is nearly density-independent near the equilibrium density $n_0$, it should be applicable not only to heavy nuclei, but also to  light nuclei. Therefore it precisely corresponds to the quenched $g_A$ observed in {\it simple} shell model in light nuclei, $g_A^L=g_A^{\rm eff}\approx 1.0$.
\item The next question is:  What does it  correspond to in what is calculated  in the high-powered microscopic calculations often heralded as ``fist-principle approach to nuclear theory," such as {\it ab initio} no-core shell models etc. Since the Gamow-Teller operator is scale-invariant as mentioned above, it has no IDD effect.  This implies that the quenching must then be entirely accounted for by many-body correlations, i.e., ``reducible diagrams" in Weinberg's classification. It is easy to understand that the $\sigma\tau$ operator responsible for the Gamow-Teller transition couples strongly to particle-hole excitations of energy  $\sim (200-300)$ MeV. But it also couples strongly to $\Delta$-hole excitations of comparable energy,  $m_\Delta-m_N \sim 300$ MeV. In both couplings, the tensor force is involved, the former in the $NN$-$NN$ channel and the latter in the $NN$-$N\Delta$ channel. Consisting of one-pion exchange and one-$\rho$
exchange, both are subject to the same density-dependent cancellation leading to the dropping tensor-force strength mentioned above. Therefore the configuration space involved in the microscopic calculations must include, not just particle-hole configurations but also $\Delta$-hole states. In 1970's, this was classified as one of several two-body exchange currents. In the modern development, this belongs to
N$^3$LO in the chiral counting. It is a part of those terms that are ``chiral-filter unprotected" as explained in Section \ref{chiral filter}. Treating it as an exchange current as is done in many of the recent works is not consistent, so cannot be trusted.
\item In the large $N_c$ limit, the nucleon and the $\Delta$ resonance are degenerate so one may consider treating them together in a generalized Fermi-liquid description in the space of $N$ and $\Delta$. Perhaps one can formulate it as a generalized Migdal interaction $g_0^\prime$ involving the three channels, $NN$, $N\Delta$ and $\Delta\Delta$. In reality, however, with the mass difference comparable to the Fermi momentum $\sim k_F$ involved, how to define  $\bar{N}$ and invoke  the large $\bar{N}$ limit to arrive at the Fermi-liquid fixed point  is problematic.  There is no satisfactory resolution of the quenched $g_A$ problem in this approach. Recall that the large $N_c$ and large $\bar{N}$ result $g_A^L=g_A^{\ast}\approx 1.0$ obtained above corresponds to having the $\Delta$-hole configurations integrated out.
\end{enumerate}
The upshot is that there is no vacuum-change-induced quenching of $g_A$ in nuclear medium.  The effective $g_A^{\rm eff}\approx 1$ observed in simple shell model is entirely due to nuclear correlations with the nuclear tensor force playing the dominant role.  We arrive at this by implementing scale symmetry -- no IDD -- and hidden local symmetry -- dropping tensor force -- in nuclear interactions. These symmetries are not visible at nuclear matter density, most likely hidden inside complex nuclear processes.
\section{Unhiding hidden symmetries in compact stars}
\subsection{Pseudo-conformal sound velocity}
We now turn to the signal for an emergent scale symmetry, as density exceeds $n_{1/2}$, with the appearance of the sound velocity approaching $v_s^2=1/3$. This velocity, one can easily see,   would come out trivially if one set the trace of energy-momentum tensor $\theta^\mu_\mu$ equal to zero. Therefore it is referred to in the literature as ``conformal sound velocity."  However $\theta_\mu^\mu$ cannot be zero in QCD (in the vacuum) because  of the  trace anomaly, which is not zero.

It was discovered in the W/H work that the sound velocity in compact stars in  EFT-$bs$HLS theory did converge to and stayed at $v_s^2=1/3$ as density exceeded $n_{1/2}$~\cite{pklmr/mlpr}.  This was a surprise since $\theta_\mu^\mu$ cannot be zero in the half-skyrmion phase. This is because the $\sigma$ mass is proportional to the dilaton condensate which is just the dilaton decay constant, which, as mentioned above, equals the chiral invariant nucleon mass $m_0\simeq (0.6-0.9)m_N$. Then how does the ``conformal sound velocity" come about? The answer is that the trace of energy momentum tensor depends on the dilaton condensate that becomes density-independent for density $n\gsim 3n_0$. Then $v_s^2=1/3$ follows provided there are no abnormal states like the Lee-Wick state. Indeed there is no indication for any of them.  Since $v_s^2=1/3$ here is not due to the vanishing of $\theta_\mu^\mu$, we refer this as ``pseudo-conformal sound velocity" and the model that leads to the appropriate EoS as ``pseudo-conformal model (PCM)."

This  prediction is quite novel,  not shared by S$\chi$EFT. In fact it is in total disagreement with what most of the theorists working in the field argue for~\cite{tews}.
\subsection{Cheshire-Cat phenomenon}
The pseudo-conformal sound velocity  brings about a remarkable thing. It is generally accepted on both phenomenological and theoretical considerations that converging to -- and remaining at -- $v_s^2=1/3$ for non-asymptotic density  is impossible {\it unless there is a change of degrees of freedom.} We are not aware of any rigorous proof for this. But if this is true, then this implies that the topology change is playing the role of a hadron-quark transition in terms of QCD variables. We identify this as a manifestation of the Cheshire Cat phenomenon discovered in 1980's~\cite{CC}.\footnote{The Cheshire-Cat phenomenon  is the continuity from hadrons to quarks in terms of the MIT bag as QCD and the skyrmion as hadron. It has also been formulated at low energy as a gauge degree of freedom.}  The $V_{lowk}$RG approach with $bs$HLS, i.e., $V_{\rm lowk}$-$bs$HLS, is exposing a pseudo-conformal symmetry, which is not in QCD proper but induced in medium. Thus the sound velocity observed in this  approach for $n \gsim  3n_0$ up to the central density in massive stars, $\sim (5-7)n_0$, is indicating an emergent scale symmetry in dense medium.
%
\subsection{Tidal deformability and $n_{1/2}$}
There is no fine-tuning involved with the PCM in contrast to many of the energy-density functional models that purport, with numerous free parameters, to fit accurately both nuclear matter and compact-star matter. The PCM has only a few parameters to play with. It does, however,  give nuclear matter and compact-star matter reasonably well, within the available error bars established. As post-dictions, there is nothing special about the PCM. What is remarkable, apart from the novel -- and unorthodox -- prediction for the pseudo-conformal sound velocity, is that the cusp structure of the tensor force, hence in the symmetry energy, has an interesting correlation with the tidal deformability (TD), i.e., the dimensionless TD $\Lambda$,  observed in the gravitational waves coming from coalescing neutron stars.  This issue was treated somewhat peripherally in the W/H project, but it became highly topical more recently due to the GW170817 event~\cite{ligo/virgo}.

The information provided by the TD is for the density regime $n\lsim 2n_0$, so it does not zero-in on the high density inside compact stars, the focus of the W/H. It is interesting however for the W/H development because the density $\sim 2n_0$ is just where the cusp structure  takes place in the symmetry energy due to the topology change. What is indicated in the dimensionless TD, $\Lambda$,  is softness in the symmetry energy with hard EoS apparently ruled out by the current bound on $\Lambda$, and hence must be relevant to the region where the symmetry energy drops as density increases, namely, below $n_{1/2}$.  This indicates that the higher $n_{1/2}$ the softer the symmetry energy $E_{sym}$ will be. Depending on the parameters, $E_{sym}$ could vanish or even become negative with increasing density if there were no topology change.   A reasonable range of $n_{1/2}$, expected from what's known up to date, is $\simeq (2-4)n_0$. This range is also suggested in models that have hadron-quark continuity -- with the quarks strongly coupled -- in the EoS.  The  GW170817 event presently gives the bound $\Lambda < 800$ for a star of $1.4M_\odot$,  a radius of (11 - 12) km and a central density $\sim 2n_0$. This bound is found to require the PCM to have $n_{1/2} > 2.0n_0$. Thus it locates where $n_{1/2}$ can be.  Now if the bound happened to go down further, then it would mean that $n_{1/2}$ must increase above $2n_0$.  It cannot go up too high, otherwise the usual star properties will go haywire. Thus the precision determination (in the future) of the bound for $\Lambda$ will allow $n_{1/2}$ to be pinned down. This is an extremely important issue for locating where the EFT-QCD changeover  expected on a general ground will set in. There is no other known means, experimental or theoretical, to determine the location of $n_{1/2}$. It is amusing that the gravitational waves could provide the means to fix the topology change density, which no up-to-date available theories or models or terrestrial experiments could do.

\section{Tribute to Gerry Brown}
We would like to point out that what transpired from the W/H project -- and what took place after the W/H that figures in the monograph~\cite{cnd3} -- goes a long way towards supporting the late Gerry Brown's long-standing approach to basic research in physics, in particular his thesis that the manifestation of chiral symmetry in nuclear interactions should be explored in the properties of hidden local symmetry and emerging scale symmetry.  In his last, unpublished, paper, he stated~\cite{gerry} that   ``Landau-Migdal fixed point interaction $G_0^\prime$ (a.k.a., `Ericson-Ericson-Lorentz-Lorenz term'), nuclear tensor force and `Brown-Rho scaling'~\cite{br91}  run the show and make all nuclear forces equal." In fact this, we believe, is what makes sense the whole picture arrived at in the W/H effort and treated in detail in the monograph. As Sherlock Holmes states in his approach to solving mysteries, ``When you have eliminated the impossible, whatever remains, however improbable, must be the truth."

\section{Acknowledgments}
We would like to acknowledge invaluable discussions and collaborations with the core members and participants of the WCU/Hanyang Program, Tom Kuo,  Hyun Kyu Lee,  Won-Gi Paeng and Sang-Jin Sin that led to what's described in this note and in the authors' monograph.
 Y.L. Ma is supported in part by National Science Foundation of China (NSFC) under grant No. 11475071, 11747308 and Seeds Funding of Jilin University.


\end{document}